\documentclass{aastex63}

\newcommand{\ztf}{ZTF\,J2055+4651}
\newcommand{\ztfold}{ZTF\,J2130+4420}
\newcommand{\kms}{\ensuremath{{\rm km}\,{\rm s}^{-1}}}
\newcommand{\degree}{$^{\circ}$ }
\newcommand{\msol}{M$_\odot$}
\newcommand{\rsol}{R$_\odot$}
\newcommand{\teff}{$T_{\rm eff}$}
\newcommand{\logg}{$\log(g)$}
\newcommand{\porb}{$P_{\rm orb}$}
\newcommand{\vrot}{$v_{\rm rot}\sin{i}$}
\usepackage{multirow}
\usepackage{xcolor}
\usepackage{tablefootnote}

\newcommand\aastex{AAS\TeX}
\newcommand\latex{La\TeX}

\received{June 1, 2019}
\revised{January 10, 2019}
\accepted{\today}
\submitjournal{ApJL}

\shorttitle{Roche lobe-filling hot subdwarf stars}
\shortauthors{Kupfer et al.}
\graphicspath{{./}{figures/}}

\begin{document}

\title{A new class of Roche lobe-filling hot subdwarf binaries}

\correspondingauthor{Thomas Kupfer}
\email{tkupfer@ucsb.edu}

\author[0000-0002-6540-1484]{Thomas Kupfer}
\affiliation{Kavli Institute for Theoretical Physics, University of California, Santa Barbara, CA 93106, USA}


\author[0000-0002-4791-6724]{Evan B. Bauer}
\affiliation{Kavli Institute for Theoretical Physics, University of California, Santa Barbara, CA 93106, USA}

\author[0000-0002-7226-836X]{Kevin B. Burdge}
\affiliation{Division of Physics, Mathematics and Astronomy, California Institute of Technology, Pasadena, CA 91125, USA}

\author[0000-0002-2626-2872]{Jan van~Roestel}
\affiliation{Division of Physics, Mathematics and Astronomy, California Institute of Technology, Pasadena, CA 91125, USA}

\author[0000-0001-8018-5348]{Eric C. Bellm}
\affiliation{DIRAC Institute, Department of Astronomy, University of Washington, 3910 15th Avenue NE, Seattle, WA 98195, USA}

\author{Jim Fuller}
\affiliation{Division of Physics, Mathematics and Astronomy, California Institute of Technology, Pasadena, CA 91125, USA}

\author[0000-0001-5941-2286]{JJ Hermes}
\affiliation{Department of Astronomy, Boston University, 725 Commonwealth Ave., Boston, MA 02215, USA}

\author[0000-0002-2498-7589]{Thomas R. Marsh}
\affiliation{Department of Physics, University of Warwick, Coventry CV4 7AL, UK}

\author{Lars Bildsten}
\affiliation{Kavli Institute for Theoretical Physics, University of California, Santa Barbara, CA 93106, USA}
\affiliation{Department of Physics, University of California, Santa Barbara, CA 93106, USA}

\author[0000-0001-5390-8563]{Shrinivas R. Kulkarni}
\affiliation{Division of Physics, Mathematics and Astronomy, California Institute of Technology, Pasadena, CA 91125, USA}

\author[0000-0002-9656-4032]{E.~S. Phinney}
\affiliation{Theoretical Astrophysics, 350-17 California Institute of Technology, Pasadena CA 91125, USA}

\author{Thomas A. Prince}
\affiliation{Division of Physics, Mathematics and Astronomy, California Institute of Technology, Pasadena, CA 91125, USA}

\author[0000-0003-4373-7777]{Paula Szkody}
\affiliation{University of Washington, Department of Astronomy, Box 351580, Seattle, WA 98195, USA}

\author[0000-0001-6747-8509]{Yuhan Yao}
\affiliation{Division of Physics, Mathematics and Astronomy, California Institute of Technology, Pasadena, CA 91125, USA}

\author[0000-0002-0465-3725]{Andreas Irrgang}
\affiliation{Dr.\ Karl Remeis-Observatory \& ECAP, Astronomical Institute, Friedrich-Alexander University Erlangen-Nuremberg (FAU), Sternwartstr.\ 7, 96049 Bamberg, Germany}

\author[0000-0001-7798-6769]{Ulrich Heber}
\affiliation{Dr.\ Karl Remeis-Observatory \& ECAP, Astronomical Institute, Friedrich-Alexander University Erlangen-Nuremberg (FAU), Sternwartstr.\ 7, 96049 Bamberg, Germany}

\author{David Schneider}
\affiliation{Dr.\ Karl Remeis-Observatory \& ECAP, Astronomical Institute, Friedrich-Alexander University Erlangen-Nuremberg (FAU), Sternwartstr.\ 7, 96049 Bamberg, Germany}

\author[0000-0003-4236-9642]{Vik S. Dhillon}
\affiliation{Department of Physics \& Astronomy, University of Sheffield, Sheffield S3 7RH, UK}
\affiliation{Instituto de Astrofísica de Canarias, Via Lactea s/n, La Laguna, E-38205 Tenerife, Spain}

\author[0000-0001-7809-1457]{Gabriel Murawski}
\affiliation{Gabriel Murawski Private Observatory (SOTES), Poland}

\author{Andrew J. Drake}
\affiliation{Division of Physics, Mathematics and Astronomy, California Institute of Technology, Pasadena, CA 91125, USA}

\author[0000-0001-5060-8733]{Dmitry A. Duev}
\affiliation{Division of Physics, Mathematics and Astronomy, California Institute of Technology, Pasadena, CA 91125, USA}

\author{Michael Feeney}
\affiliation{Caltech Optical Observatories, California Institute of Technology, Pasadena, CA 91125, USA}

\author[0000-0002-3168-0139]{Matthew J. Graham}
\affiliation{Division of Physics, Mathematics and Astronomy, California Institute of Technology, Pasadena, CA 91125, USA}

\author[0000-0003-2451-5482]{Russ R. Laher}
\affiliation{IPAC, California Institute of Technology, 1200 E. California Blvd, Pasadena, CA 91125, USA}

\author[0000-0001-7221-855X]{S. P. Littlefair}
\affiliation{Department of Physics \& Astronomy, University of Sheffield, Sheffield S3 7RH, UK}

\author[0000-0003-2242-0244]{A.~A.~Mahabal}
\affiliation{Division of Physics, Mathematics and Astronomy, California Institute of Technology, Pasadena, CA 91125, USA}
\affiliation{Center for Data Driven Discovery, California Institute of Technology, Pasadena, CA 91125, USA}

\author[0000-0002-8532-9395]{Frank J. Masci}
\affiliation{IPAC, California Institute of Technology, 1200 E. California Blvd, Pasadena, CA 91125, USA}

\author{Michael Porter}
\affiliation{Caltech Optical Observatories, California Institute of Technology, Pasadena, CA 91125, USA}

\author{Dan Reiley}
\affiliation{Caltech Optical Observatories, California Institute of Technology, Pasadena, CA 91125, USA}

\author{Hector Rodriguez}
\affiliation{Caltech Optical Observatories, California Institute of Technology, Pasadena, CA 91125, USA}

\author[0000-0001-7648-4142]{Ben Rusholme}
\affiliation{IPAC, California Institute of Technology, 1200 E. California Blvd, Pasadena, CA 91125, USA}

\author[0000-0003-4401-0430]{David L. Shupe}
\affiliation{IPAC, California Institute of Technology, 1200 E. California Blvd, Pasadena, CA 91125, USA}

\author[0000-0001-6753-1488]{Maayane T. Soumagnac}
\affiliation{Lawrence Berkeley National Laboratory, 1 Cyclotron Road, Berkeley, CA 94720, USA}
\affiliation{Department of Particle Physics and Astrophysics, Weizmann Institute of Science, Rehovot 76100, Israel}



\begin{abstract}
We present the discovery of the second binary with a Roche lobe-filling hot subdwarf transferring mass to a white dwarf (WD) companion. This 56 minute binary was discovered using data from the Zwicky Transient Facility. Spectroscopic observations reveal an He-sdOB star with an effective temperature of \teff$=33,700\pm1000$\,K and a surface gravity of \logg$=5.54\pm0.11$. The GTC+HiPERCAM light curve is dominated by the ellipsoidal deformation of the He-sdOB star and shows an eclipse of the He-sdOB by an accretion disk as well as a weak eclipse of the WD. We infer a He-sdOB mass of $M_{\rm sdOB}=0.41\pm0.04$\,\msol\, and a WD mass of $M_{\rm WD}=0.68\pm0.05$\,\msol. The weak eclipses imply a WD black-body temperature of $63,000\pm10,000$\,K and a radius $R_{\rm WD}=0.0148\pm0.0020$\,\rsol\, as expected for a WD of such high temperature.

The He-sdOB star is likely undergoing hydrogen shell burning and will continue transferring mass for $\approx1$\,Myrs at a rate of $10^{-9}\ M_\odot\, {\rm yr}^{-1}$ which is consistent with the high WD temperature. The hot subdwarf will then turn into a WD and the system will merge in $\approx30$\,Myrs. We suggest that Galactic reddening could bias discoveries towards preferentially finding Roche lobe-filling systems during the short-lived shell burning phase. Studies using reddening corrected samples should reveal a large population of helium core-burning hot subdwarfs with \teff$\approx25,000$\,K in binaries of 60-90 minutes with WDs. Though not yet in contact, these binaries would eventually come into contact through gravitational wave emission and explode as a sub-luminous thermonuclear supernova or evolve into a massive single WD. 

\end{abstract}



\section{Introduction} \label{sec:intro}
Subdwarf B stars (sdBs) are stars of spectral type B with luminosities below the main sequence. The formation mechanism and evolution of sdBs are still debated, although most sdBs are likely helium (He)--burning stars with masses $\approx0.5$\,\msol\,and thin hydrogen envelopes (\citealt{heb86,heb09,heb16}). A large fraction are found in binary systems \citep{nap04a,max01} and the most compact ones have periods $\lesssim\,1$\,hr (e.g. \citealt{ven12,gei13,kup17,kup17a,kup20}). Systems with orbital periods $\lesssim$\,2\,hrs at the exit of the last common envelope will overflow their Roche Lobe while the hot subdwarf is still burning helium (e.g. \citealt{sav86,tut89,tut90,ibe91,yun08,pie14,bro15}). 

So far only a few hot subdwarf binaries have been identified that will start mass transfer while the hot subdwarf is still burning helium in its core. The first discovered system is CD--30$^{\circ}$11223 which has an orbital period of \porb=70.5\,min, and will start mass transfer in $\approx\,40$\,Myrs \citep{ven12,gei13}.

Most recently \citet{kup20} discovered \ztfold, the first system where the sdOB fills its Roche lobe and has started mass transfer to its white dwarf (WD) companion. In this article, we report the discovery of ZTF\,J205515.98+465106.5 (hereafter \ztf), the second confirmed member of Roche Lobe-filling hot subdwarf stars in compact binaries. \ztf\, was discovered in a period search using the conditional entropy period-finding algorithm \citep{gra13} of all Pan-STARRS objects with colors of $g-r<0.2$ and $r-i<0.2$, and more than 50 ZTF measurements at the time, as described in \citet{bur19} and \citet{cou20}. Conditional entropy is an information theory-based algorithm that seeks to find the period that produces the most ordered (lowest entropy) arrangement of data points in the phase folded light curve whilst also accounting for the phase space coverage of the data. Independently, on 2019-07-19, Gabriel Murawski reported the object to the International Variable Star Index (VSX) \footnote{https://www.aavso.org/vsx/index.php?view=detail.top\&oid=838260}. Also, \citet{riv20} report the absence of X-ray emission from 3.7\,ks {\it Swift} observations and found a limit for the X-ray flux of $<1\times10^{-13}$\,erg\,s$^{-1}$\,cm$^{-2}$ leading to a limit of $<6\times10^{31}$\,erg\,s$^{-1}$ at a distance of 2170\,pc \citep{bai18}. We present the system properties, compare them to \ztfold\, and discuss possible implications for this new class of Roche lobe-filling hot subdwarf binaries.


\section{Observations} \label{sec:observations}
As part of the Zwicky Transient Facility (ZTF), the Palomar 48-inch (P48) telescope images the sky every clear night. \ztf\, was discovered as part of a dedicated high-cadence ZTF survey at low Galactic latitudes \citep{bel19,gra19}, which either observed one field or alternated between two adjacent fields continuously for $\approx$1.5-3\,hours on two to three consecutive nights in the ZTF-$r$ band \citep{bel19a}. The high-cadence data was complemented by data from the ZTF public survey which was made available after data release 2 on Dec 11, 2019. Image processing of ZTF data is described in full detail in \citet{mas19}. The ZTF lightcurve of \ztf\, consists of 224 observations in ZTF-$g$ and 496 observations in ZTF-$r$.

Additionally on 2019-09-05, \ztf\, was observed with the Gran Telescopio Canarias (GTC) using HiPERCAM, which is a five-beam imager equipped with frame-transfer CCDs allowing the simultaneous acquisition of $u_s$, $g_s$, $r_s$, $i_s$ and $z_s$\footnote{HiPERCAM uses high-throughput versions of the SDSS filters known as Super-SDSS filters, and we denote these filters $u_s$, $g_s$, $r_s$, etc.} band images at  a rate of up to 1000 frames per second \citep{dhi16, dhi18}. \ztf was observed at a 3\,s cadence with a dead time of 10\,ms for 1130 frames with HiPERCAM. The run lasted  1\,hr, covering a little more than one 56\,min binary orbit. The cadence in $u_s$ and $z_s$ alone was a factor of 3 slower than the other bands to optimise signal-to-noise. The data were reduced using the dedicated HiPERCAM pipeline\footnote{https://github.com/HiPERCAM/}, including debiasing and flat-fielding. Differential photometry was also performed. Due to technical problems the $i_s$ band light curve could not be used for the analysis.

On 2019-07-30 a sequence of 5 optical spectra was obtained with the Shane 
telescope and the KAST spectrograph at Lick observatory using a low resolution mode ($R\approx1200$). The spectra were taken consecutively and had an exposure time of 12\,min. Data reduction was performed with the standard \texttt{IRAF} packages. Although 12\,min covers 1/5 of the orbit we detected large radial velocity shifts between the spectra, confirming the binary nature. 

Additionally, on 2019-09-24 and 2019-09-25 phase-resolved spectroscopy of \ztf\, was obtained using the Gemini North Telescope, using GMOS in long slit mode with the B600 grating ($R\approx1400$). We obtained a total of 64 spectra with an exposure time of 3\,min per spectrum. CuAr arc exposures were taken at the position of the target before and after each observing sequence to account for telescope flexure. The data were reduced using the \texttt{IRAF} package for GMOS.

\section{Period finding and spectroscopic analysis}\label{sec:datanalysis}
As we followed the same procedure as described in \citet{kup20}, we only briefly describe the major steps here. To derive the orbital period of the system, we use the ZTF light curve with its multi-month baseline in combination with the HiPERCAM light curve. The \texttt{Gatspy}\footnote{http://dx.doi.org/10.5281/zenodo.14833}\citep{van15, van15a} module that utilizes the Lomb-Scargle periodogram \citep{Lom76,sca82} was used for period finding, yielding \porb$=56.34785(26)$\,min. Using the parameter from the best light curve fit (see Sect.\,\ref{sec:system}) and only fitting the zero point of the ephemeris ($T_{o}$) to the HiPERCAM light curves, we find an ephemeris of:

\begin{equation}\label{equ:ephem}
T_{o} ({\rm BMJD}) = 58731.9639(90) + 0.03913045(18)E, 
\end{equation}

\noindent  
where $E$ corresponds to the epoch.

\begin{figure}
\begin{center}
\includegraphics[width=0.98\textwidth]{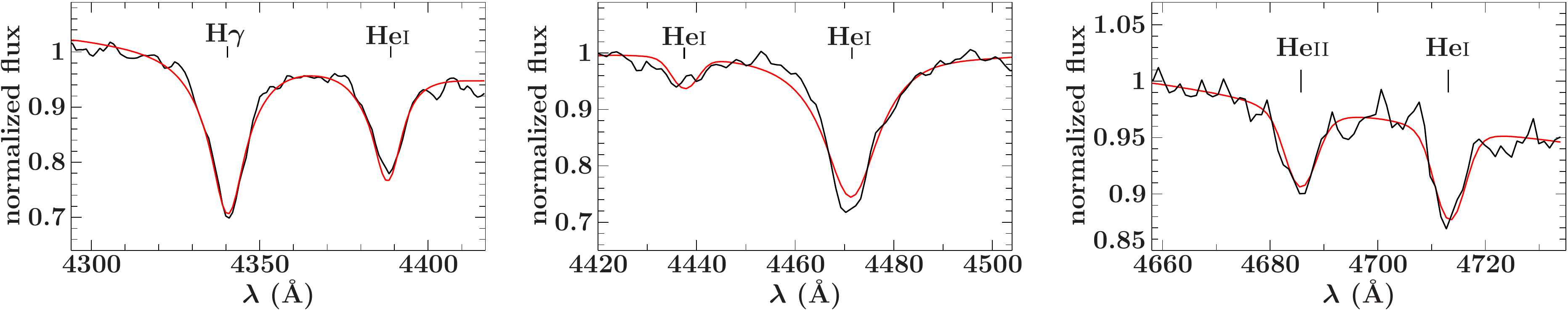}
\includegraphics[width=0.98\textwidth]{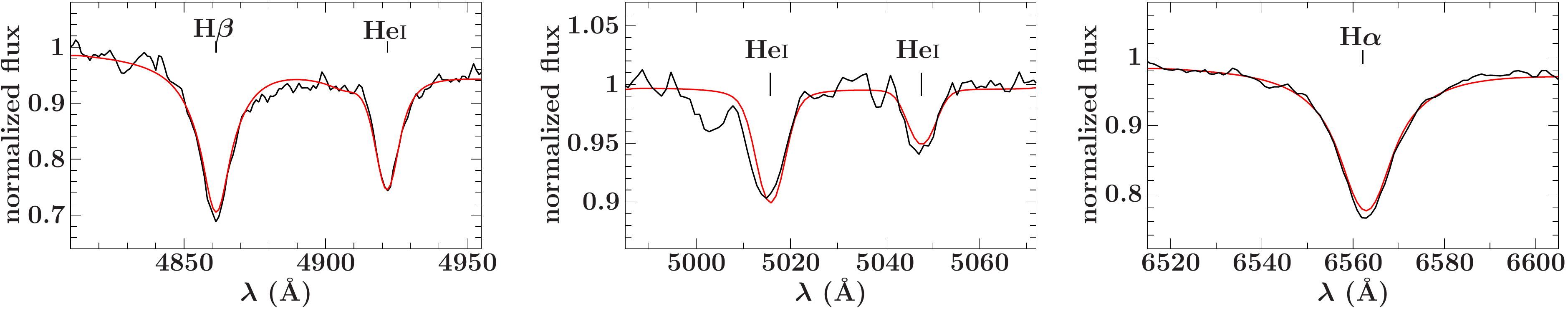}
\end{center}
\caption{Fit of synthetic NLTE models to the hydrogen Balmer, as well as neutral and ionized helium lines of the co-added Gemini spectrum. The black solid line corresponds to the spectrum and red solid line to the fit.}
\label{fig:ztf2055_spec}
\end{figure}

We find spectral features of hydrogen, neutral (He\,{\sc i}) and ionized (He\,{\sc ii}) helium lines, typical for sdOB stars. By fitting the rest-wavelength corrected average Gemini spectrum with metal-free NLTE model spectra \citep{str07}, we determine \teff$=33,700\pm1000$\,K, \logg$=5.54\pm0.11$, $\log{y}=\log[{n({\rm He})/n({\rm H})}]=0.06\pm0.05$ and \vrot$=201\pm30$\,\kms. Additionally, we measured atmospheric parameters using NLTE models computed with \texttt{TLUSTY} \citep{hub95} which include typical hot subdwarf abundances of nitrogen, carbon and oxygen \citep{nas18}. We find fully consistent results with our metal-free NLTE models. Figure \,\ref{fig:ztf2055_spec} shows the fit to the Gemini spectrum. The occurrence of both He\,{\sc i} and He\,{\sc ii} as well as the increased helium abundance $\log{y}>0$ classifies the hot subdwarf as an He-sdOB star (see \citealt{heb16}). 

\begin{figure}
\begin{center}
\includegraphics[width=0.68\textwidth]{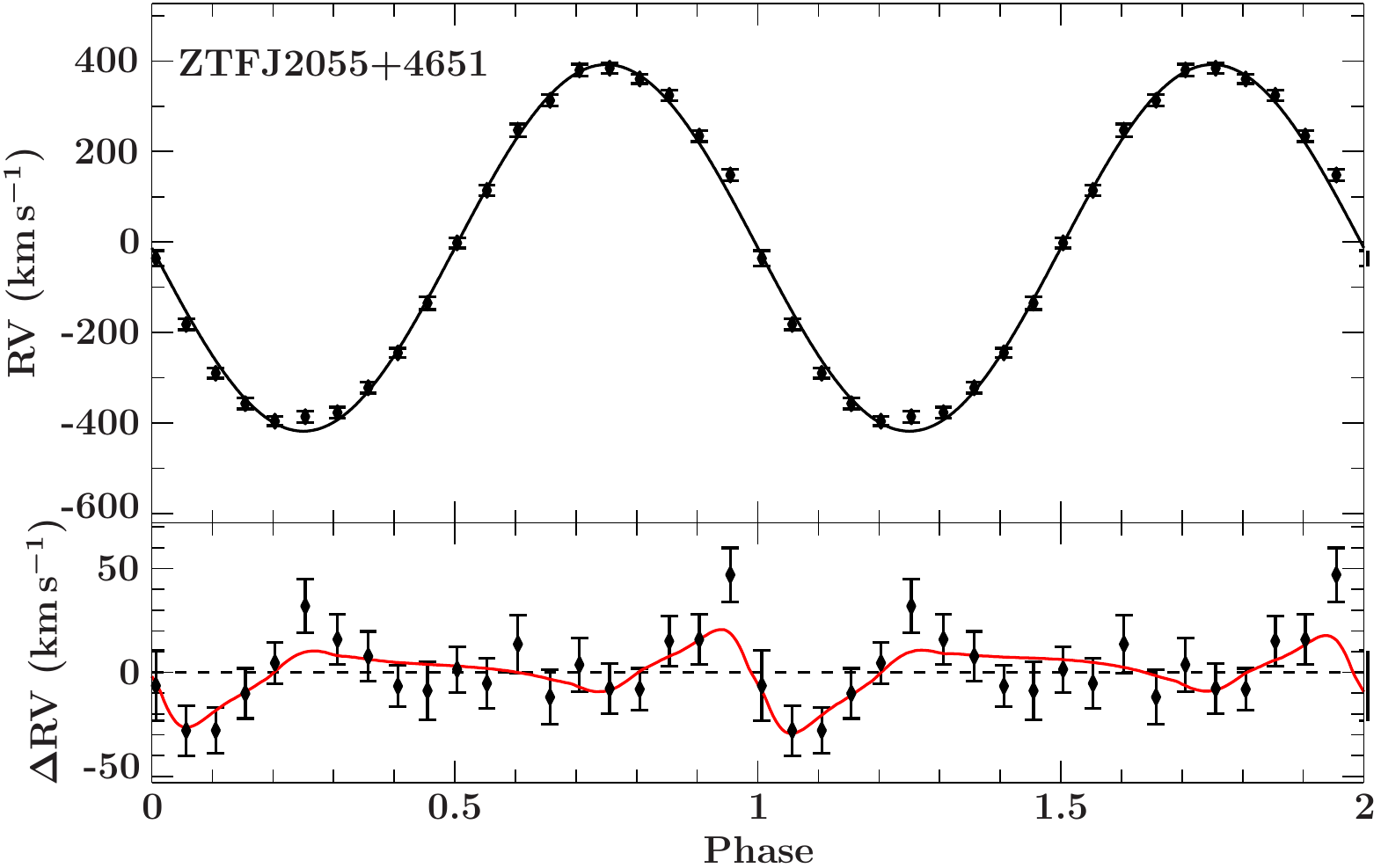}
\end{center}
\caption{Measured radial velocity versus orbital phase for \ztf. The RV data are phase folded with the orbital period and  plotted twice for better visualization. The residuals are plotted below. The strong deviation from a pure-sine curve around phase 0 (1) can be explained by the Rossiter-McLaughlin effect occurring when the accretion disk eclipses the rapidly rotating sdOB. The red curve shows the predicted residuals for the Rossiter-McLaughlin effect from our best fitting model (see Sect. \,\ref{sec:system}).} 
\label{fig:rv_curve1}

\end{figure}

We folded the individual Gemini spectra on the ephemeris shown in Equation (\ref{equ:ephem}) into 20 phase-bins and co-added individual spectra observed at the same binary phase. This leads to a signal-to-noise per phase-bin of $\approx$30. We used the \texttt{FITSB2} routine \citep{nap04a,gei11a} to  then measure radial velocities. Assuming circular orbits, a sine curve was fitted to the folded radial velocity (RV) data points (Fig.\,\ref{fig:rv_curve1}) excluding the data points around Phase $0.8-0.2$. We find a velocity semi-amplitude $K=404.0\pm11.0$\,\kms.
Similar to ZTF\,J2130, the velocity deviates significantly from a pure sine-curve around the phase when the sdOB is furthest away from the observer, which can be explained with the Rossiter-McLaughlin effect. The red curve in Fig.\,\ref{fig:rv_curve1} show the residuals predicted from the Rossiter-McLaughlin effect calculated from our best fitting light curve model.

\begin{table}
\centering
\caption{Measured and derived parameters for \ztf.}
\begin{tabular}{lll}
\hline\hline
Right ascension$^a$ & RA [hrs]  & 20:55:15.98 \\
Declination$^a$  & Dec $[^\circ]$  & +46:51:06.46 \\
Magnitude$^b$ & $g$ [mag] & 17.72$\pm$0.04 \\
Parallax$^a$  & $\varpi$    [mas] & 0.432$\pm$0.098 \\ 
Distance  &  $d$ [kpc] &  $2.3^{+0.7}_{-0.4}$  \\
reddening$^c$ & $E(g-r)$ [mag]   &  $0.46\pm0.03$ \\
Absolute Magnitude  &  \multirow{2}{*}{$M_{\rm g}$ [mag]}   & \multirow{2}{*}{$4.4\pm0.6$}  \\
(reddening corrected)$^d$ &   &  \\
Proper motion (RA)$^a$ &   $\mu_\alpha$cos$(\delta)$  [mas\,yr$^{-1}$]   &  $-3.33\pm0.17$ \\
Proper motion (Dec)$^a$ &   $\mu_\delta$  [mas\,yr$^{-1}$]   &  $-5.160\pm0.167$  \\
\hline
\multicolumn{3}{l}{Atmospheric and orbital parameters of the sdOB from spectroscopic fits}  \\ 
Effective temperature & \teff\,[K] & $33,700\pm1000$ \\
Surface gravity   & \logg  & $5.54\pm0.11$  \\
Helium abundance & $\log{y}$  & $0.06\pm0.05$ \\
Projected rotational velocity & \vrot\,[\kms] &  $201\pm30$  \\
RV semi-amplitude (sdOB) & $K$ [\kms] & $404.0\pm11.0$ \\
System velocity & $\gamma$\,[\kms] & $-20.0\pm6.0$ \\ 
Binary mass function & $f_{\rm m}$ [\msol] & $0.267\pm0.022$  \\
\hline
\multicolumn{3}{l}{Derived parameters from light curve fit combined with spectroscopic results} \\
Ephemeris zero point &  $T_0$ [MBJD]  &  $58731.9639(90)$ \\
Orbital period & \porb\,[min]  & $56.34785(26)$  \\
Mass ratio  &  $q = \frac{M_{\rm sdOB}}{M_{\rm WD}}$  & $0.60\pm0.03$  \\
sdOB mass &  $M_{\rm sdOB}$ [\msol] & $0.41\pm0.04$ \\ 
sdOB radius & $R_{\rm sdOB}$ [R$_{\odot}$] &  $0.17\pm0.01$ \\ 
WD mass &  $M_{\rm WD}$ [\msol] & $0.68\pm0.05$ \\
WD radius &  $R_{\rm WD}$ [\msol] & $0.0148\pm0.0020$ \\
Orbital inclination & $i\,[^\circ$] &  $83.4\pm1.0$  \\
Separation  & $a$ [R$_{\odot}$]   &  $0.50\pm0.01$ \\
\hline
\multicolumn{3}{l}{Derived parameters from spectrophotometry and parallax$^e$} \\
Color excess & $E(44-55)$ & $0.559\pm0.021$\,mag \\
Angular diameter & $\log(\Theta\,\mathrm{(rad)})$ & $-11.479\pm0.011$ \\
sdOB radius & $R = \Theta/(2\varpi)$ [$R_\odot$] & $0.17^{+0.05}_{-0.04}$ \\
sdOB mass & $M = g R^2/G$ [\msol] & $0.37^{+0.29}_{-0.14}$ \\
Luminosity & $L/L_\odot = (R/R_\odot)^2(T_\mathrm{eff}/T_{\mathrm{eff},\odot})^4$ & $34^{+24}_{-12}$ \\
\hline
\end{tabular}
\begin{flushleft}
$^a$ taken from {\it Gaia} DR2, epoch 2015.5 \citep{gai16,gai18} \\
$^b$ taken from PanSTARRS DR1 \citep{cham16} \\
$^c$ taken from 3D extinction maps \citep{gre19} \\ 
$^d$ reddening corrected with 3D extinction maps from \citep{gre19} \\
$^e$ the given uncertainties are single-parameter $1\sigma$ confidence intervals based on $\chi^2$ statistics. A generic excess noise of $0.026$\,mag has been added in quadrature to all photometric measurements to achieve a reduced $\chi^2$ of unity at the best fit.\\
\label{tab:system}
\end{flushleft}
\end{table}

\begin{figure*}
\begin{center}
\includegraphics[width=0.45\textwidth]{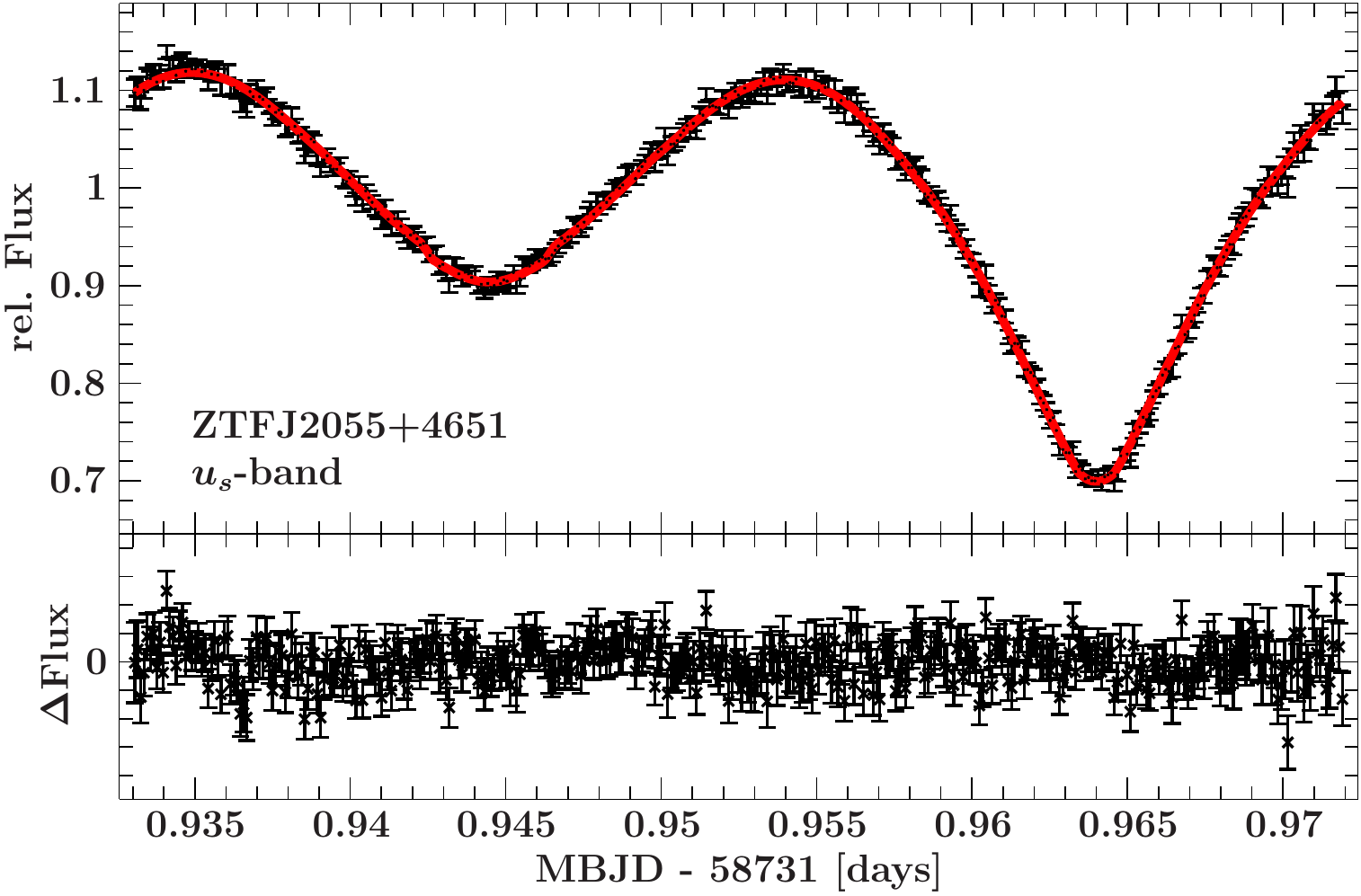}
\includegraphics[width=0.45\textwidth]{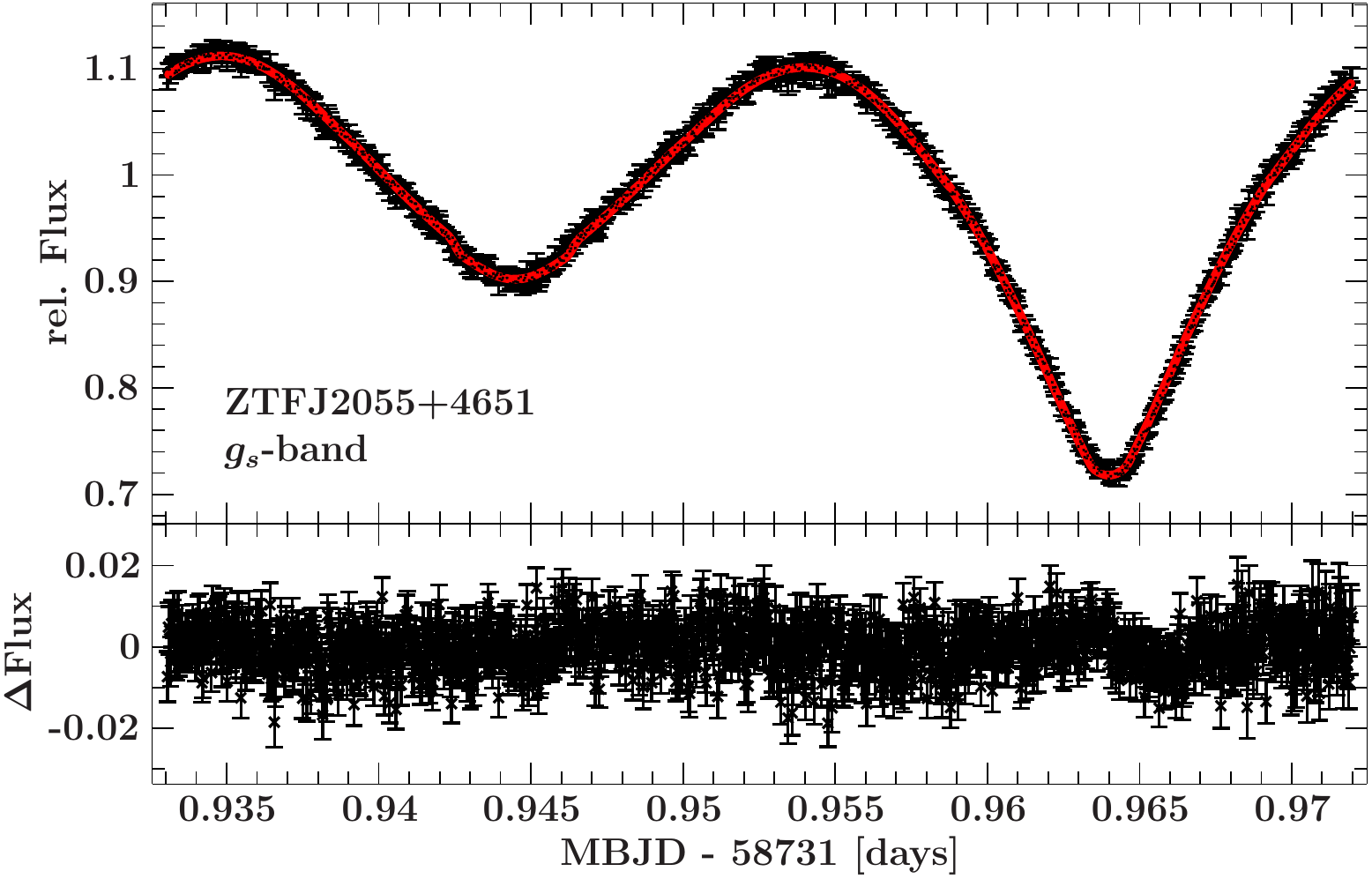}
\includegraphics[width=0.45\textwidth]{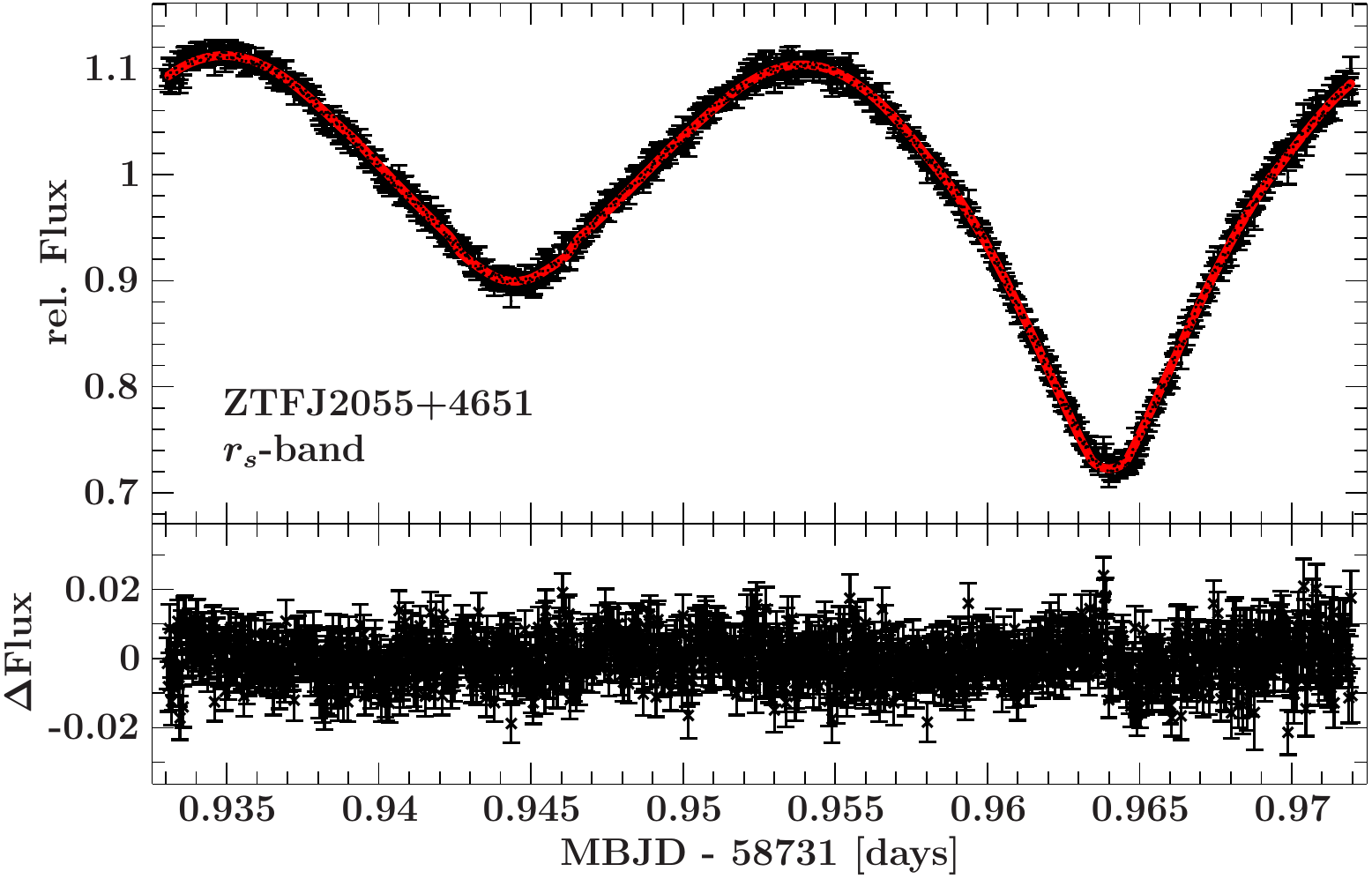}
\includegraphics[width=0.45\textwidth]{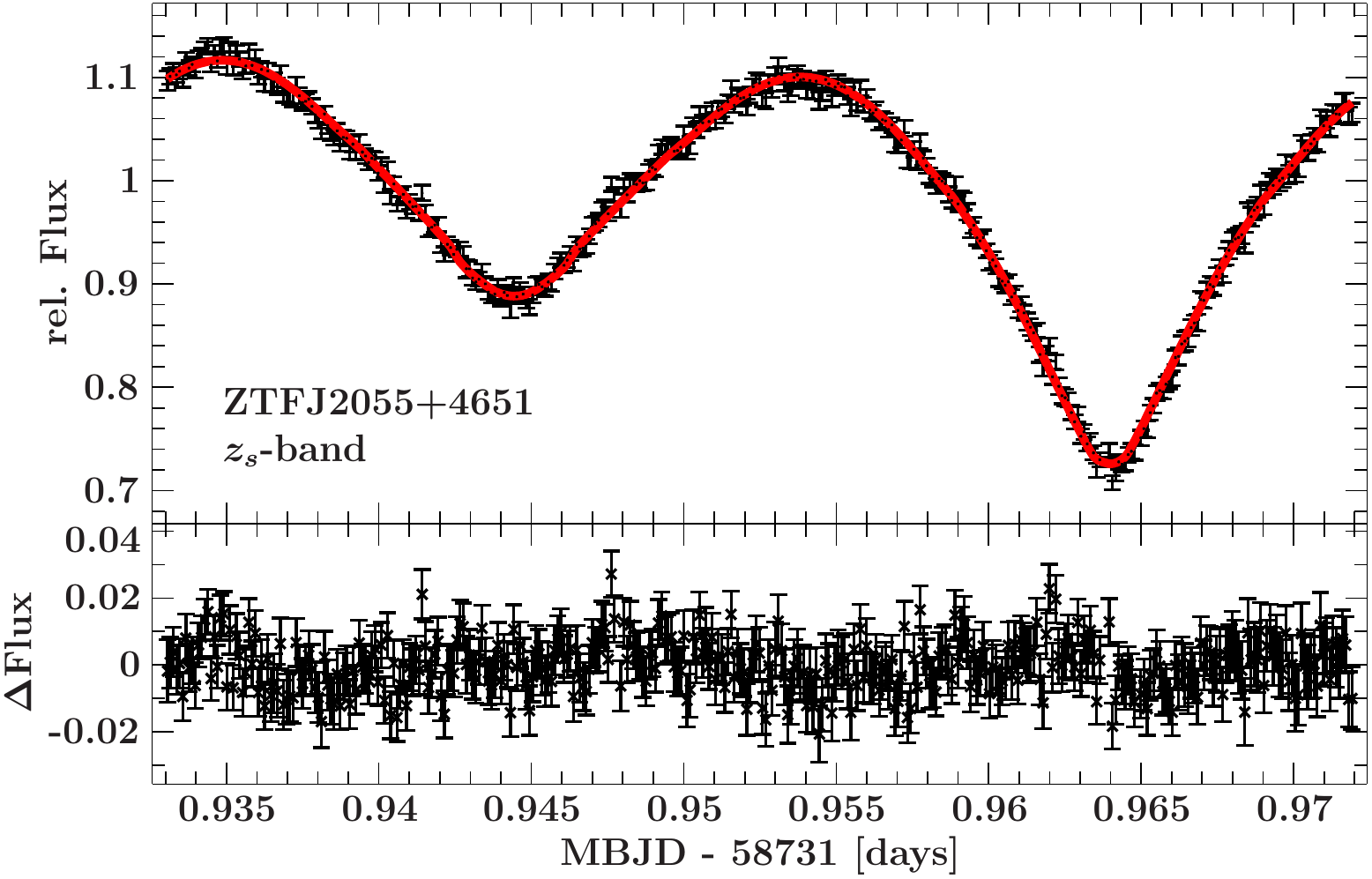}
\includegraphics[width=0.45\textwidth]{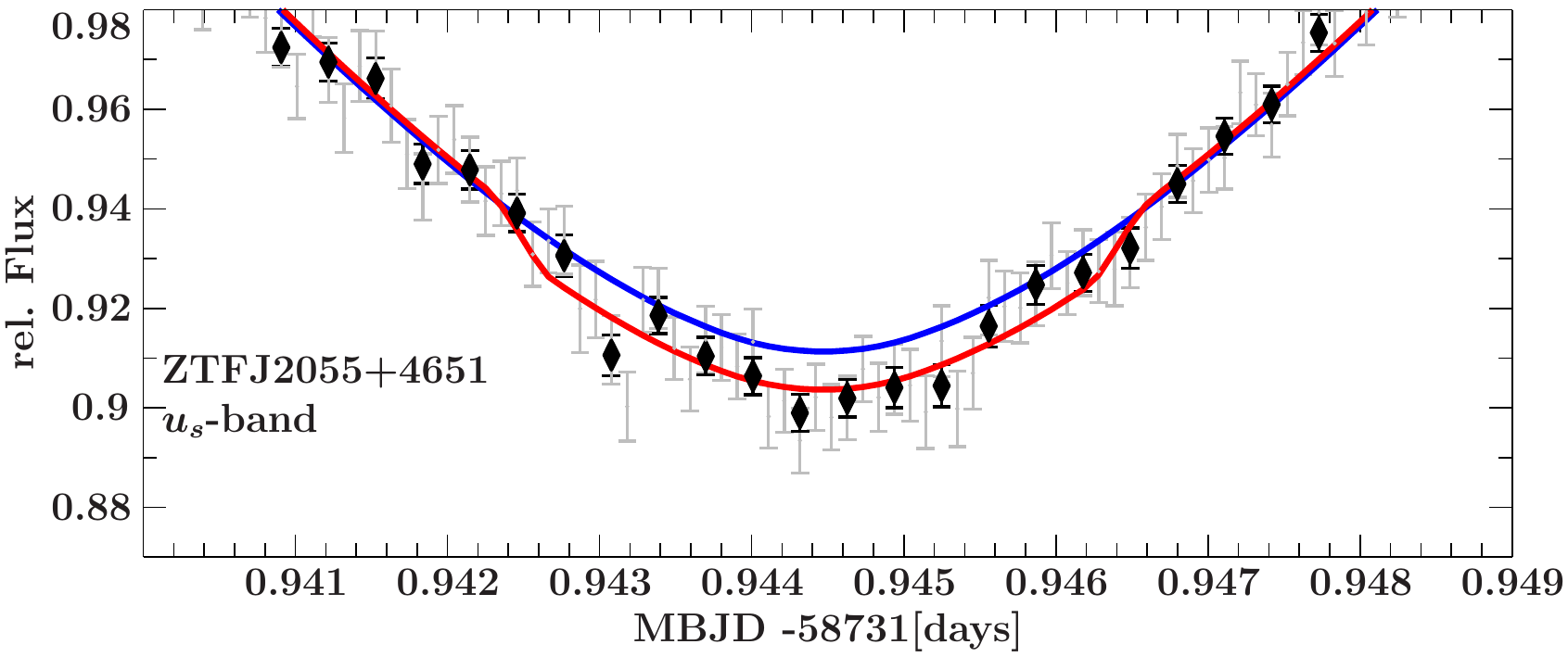}
\includegraphics[width=0.45\textwidth]{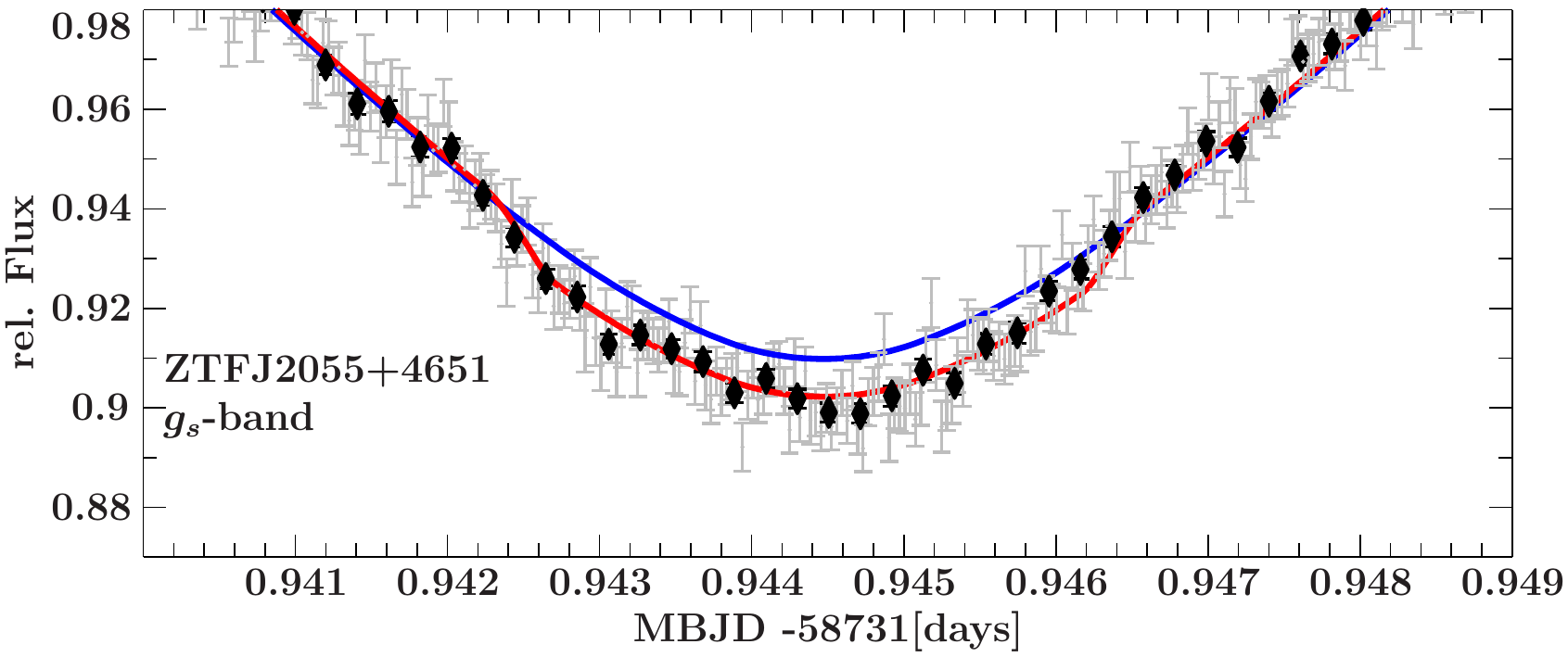}
\includegraphics[width=0.45\textwidth]{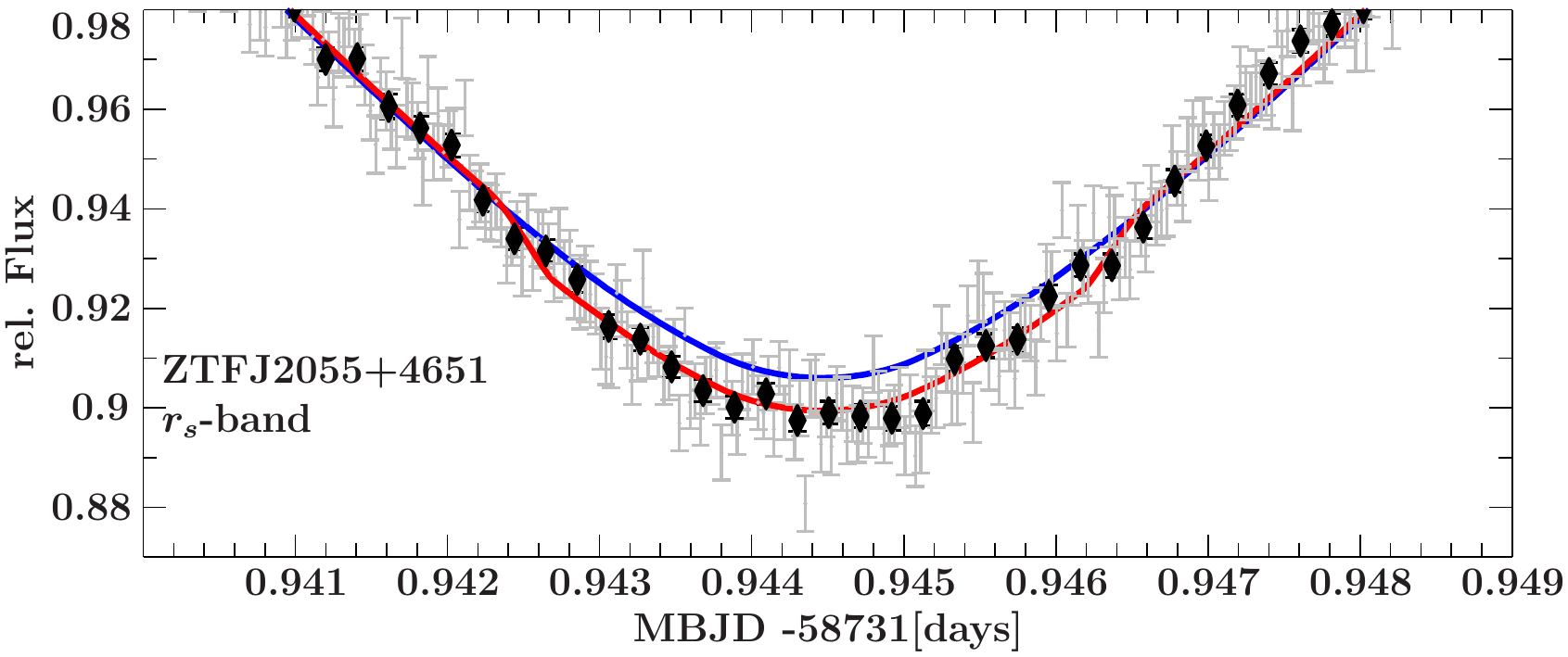}
\includegraphics[width=0.45\textwidth]{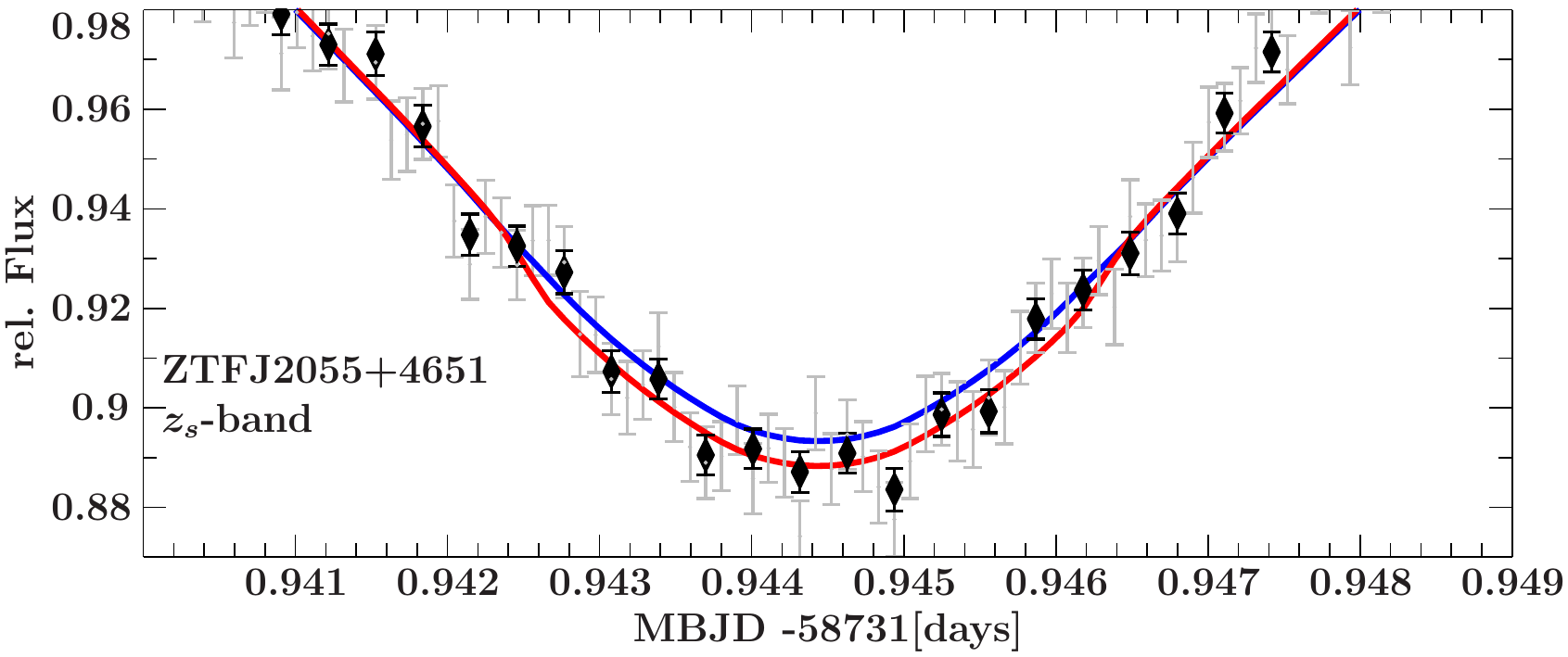}
\end{center}
\caption{HiPERCAM light curves un-binned (grey) and binned (black) shown together with the \texttt{LCURVE} fits (red) observed optical Super-SDSS bandpasses. The deeper dip corresponds to an eclipse of the He-sdOB by the accretion disc whereas the shallow dip originates in the ellipsoidal deformation of the sdOB star. The lower four panels show the region when the hot WD is being eclipsed by the He-sdOB. The blue solid curve marks the same model without eclipses of the hot WD.}
\label{fig:light_fits_2055}
\end{figure*}


\section{Light curve analysis and system parameter}\label{sec:system}
The \texttt{LCURVE} code was used to perform the light curve analysis of the HiPERCAM $u_s$, $g_s$, $r_s$ and  $z_s$ light curves \citep{cop10}. We assume a Roche lobe-filling sdOB star, a hot WD, and an irradiated disk around the WD. The passband specific beaming parameter $B$ ($F_\lambda = F_{0,\lambda} \lbrack 1 - B \frac{v_r}{c}\rbrack$, see \citealt{blo11}) was calculated following the approximation from \citet{loe03}. We use $B=1.7$, $1.5$, $1.37$ and $1.25$ for $u_s$, $g_s$, $r_s$ and  $z_s$ respectively. The passband specific gravity-darkening ($\beta$) and limb-darkening ($a_\mathrm{1}$, $a_\mathrm{2}$, $a_\mathrm{3}$, $a_\mathrm{4}$) were fixed to the theoretical values from \citet{cla11} for \teff$=35,000$\,K and \logg$=5.00$ and solar metallicity. We used $\mathrm{\beta}=0.54$, $\mathrm{a_1}=0.93$,  $\mathrm{a_2}=-0.98$, $\mathrm{a_3}=0.85$, and $\mathrm{a_4}=-0.29$ for $u_s$, $\mathrm{\beta}=0.42$, $\mathrm{a_1}=0.96$,  $\mathrm{a_2}=-1.05$, $\mathrm{a_3}=0.87$, and $\mathrm{a_4}=-0.28$ for $g_s$, $\mathrm{\beta}=0.41$, $\mathrm{a_1}=0.90$,  $\mathrm{a_2}=-1.11$, $\mathrm{a_3}=0.94$, and $\mathrm{a_4}=-0.31$ for $r_s$, and $\mathrm{\beta}=0.39$, $\mathrm{a_1}=0.83$,  $\mathrm{a_2}=-1.20$, $\mathrm{a_3}=1.09$, and $\mathrm{a_4}=-0.38$ for $z_s$. We also tried to set the limb- and gravity-darkening parameters according to the tables of \cite{cla20} for hydrogen rich atmospheres and found no significant difference in the final results. In addition, to account for any residual airmass effect we added a first order polynomial. The best value of $\chi^2$ for this model was around 3500 for 3012 points which includes also a weak eclipse of the hot WD. Although the eclipse is weak ($\leq1$\,\%), the $\chi^2$ for the non-eclipsing solution is 3900 which is statistically significantly worse compared to the solution with the weak eclipse. For the final model we normalized the errors in the data to account for the small additional residuals and obtain a reduced $\chi^2\approx1$. To determine the uncertainties in the parameters we combine \texttt{LCURVE} with \texttt{emcee} \citep{for13}, an implementation of an MCMC sampler that uses a number of parallel chains to explore the solution space. We use 256 chains and let them run until the chains stabilized to a solution, which took approximately 1500 generations. Figure~\ref{fig:light_fits_2055} shows the best fit for each HiPERCAM band.

Although \ztf\, is a single-lined binary, we can still constrain the masses and radii of the two stars by combining the light curve modeling with the spectroscopic fitting. Parameters derived in this way by a simultaneous fit to the HiPERCAM $u_s$, $g_s$, $r_s$ and $z_s$ light curves are summarized in Table \ref{tab:system}.

Our solution converges on a mass ratio of $q$ = $M_1/M_2$ = $0.60\pm0.03$, with individual masses of $M_{\rm He-sdOB} = 0.41\pm0.04$\,M$_{\odot}$ and $M_{\rm WD} = 0.68\pm0.05$\,M$_{\odot}$. The inclination is $i=83.4\pm1.0$\,\degree\, and the He-sdOB is Roche lobe-filling and has a volumetric-corrected radius of $R_{\rm sdOB}=0.17\pm0.01$\rsol\, (Tab.\,\ref{tab:system}). We find that the He-sdOB would have a projected rotational velocity \vrot$=215\pm6$\,\kms\ if synchronized to the orbit, consistent with the observed \vrot$=201\pm30$\,\kms. We detect a weak eclipse of the hot WD which constrains its properties (Fig.~\ref{fig:light_fits_2055}). We derive a black-body temperature of $63,000\pm10,000$\,K for the WD, with a radius of $0.0148\pm0.0020$\,\rsol\, which is $10$-$50$\,\% larger than a fully degenerate WD. This is consistent with predictions for hot WDs as shown in \citet{rom19} who predict $25$\,\% increased radius for a carbon-oxygen WD with \teff$=60,000$\,K and a hydrogen layer of $M=10^{-4}$\,\msol.

\begin{figure*}
\centering
\includegraphics[width=0.65\textwidth]{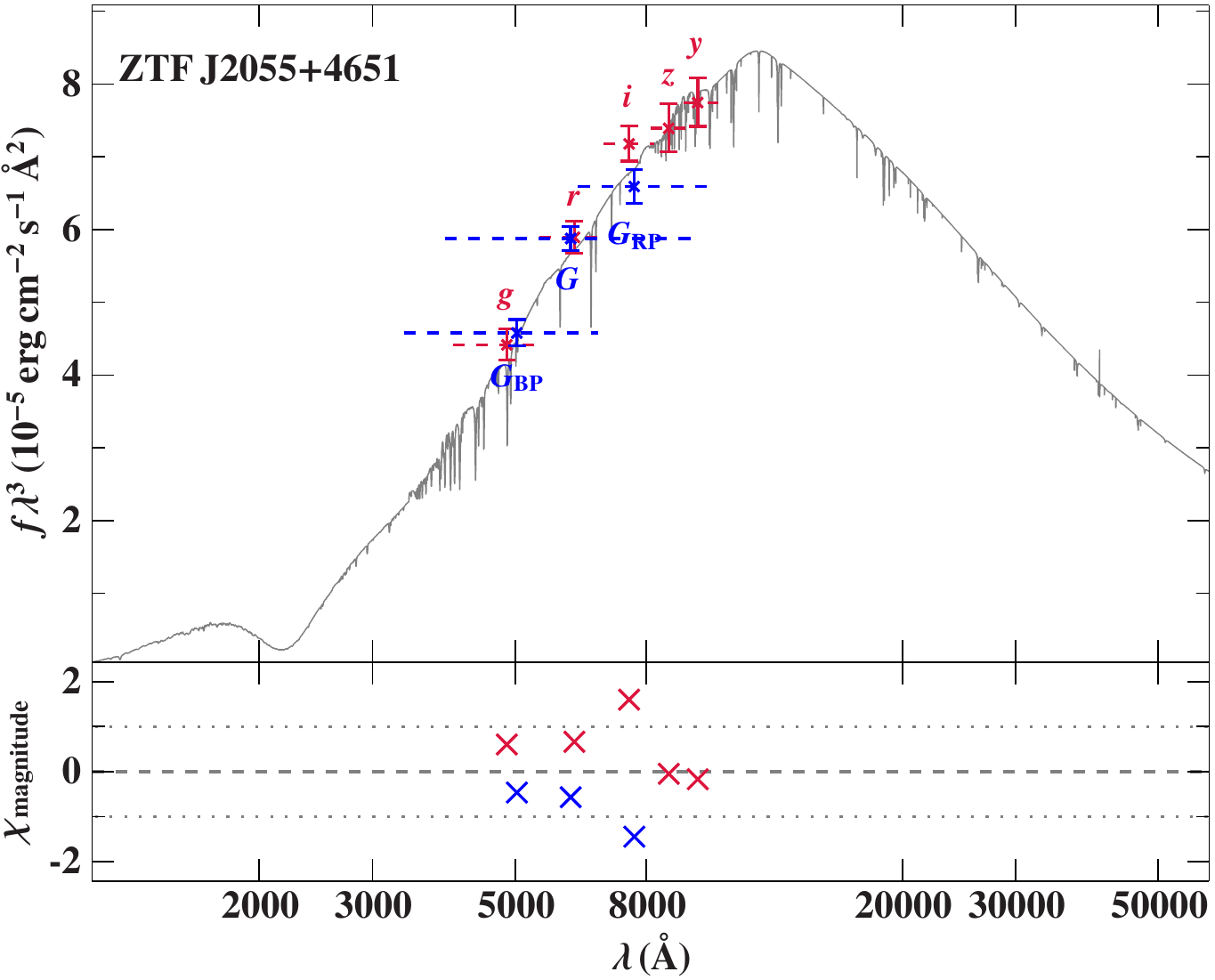}
\caption{\label{fig:photometry_sed}Comparison of synthetic and observed photometry: The \textit{top panel} shows the SED. The colored data points are filter-averaged fluxes which were converted from observed magnitudes (the respective filter widths are indicated by the dashed horizontal lines), while the gray solid line represents the best-fitting model, i.e., it is based on the parameters from Table~\ref{tab:system}, degraded to a spectral resolution of 6\,{\tiny\AA}. The flux is multiplied with the wavelength to the power of three to reduce the steep slope of the SED. The panel at the \textit{bottom} shows the residuals $\chi$, i.e., the difference between synthetic and observed magnitudes divided by the corresponding uncertainties. Only the angular diameter $\Theta$ and the color excess $E(B-V)$ are free fitting parameters. Pan-STARRS1 magnitudes \cite{cham16} are shown in red and {\it Gaia} magnitudes \citep{2018A&A...616A...4E} with corrections and calibrations from \cite{2018A&A...619A.180M} in blue.}
\end{figure*}

The spectral energy distribution (SED) in combination with the {\it Gaia} parallax allows for an independent estimate of the stellar parameters of the hot subdwarf. To this end, we first construct the observed SED from {\it Gaia} \citep{2018A&A...616A...4E} and PanSTARRS photometry \citep{cham16} \footnote{Based on averages of 223 epochs of good Gaia observations and of 80 epochs of PanSTARRS observations}. Using the atmospheric parameters derived by spectroscopy, a synthetic SED is computed with an updated version of {\sc Atlas12} \citep{1996ASPC..108..160K,2018A&A...615L...5I}, which is then matched to the observed SED by fitting the angular diameter $\Theta$ as a distance scaling factor and $E(44-55)$ as interstellar reddening parameter \citep[see][for details]{2018OAst...27...35H}. The interstellar extinction law by \citet{2019ApJ...886..108F} is used. The results are summarized in Table~\ref{tab:system} and visualized in Fig.~\ref{fig:photometry_sed}. As expected for a star in the Galactic disk, reddening is large ($E(44-55)=0.559\pm0.021$\,mag).

In combination with the {\it Gaia} parallax $\varpi$ and the atmospheric parameters (\teff\ and \logg) the angular diameter allows us to determine the stellar parameters ($R_{\rm sdOB}$, $M_{\rm sdOB}$, and luminosity $L_{\rm sdOB}$). Results are listed Table~\ref{tab:system}. The mass derived for the hot subdwarf agrees well with that derived from the light curve, although the uncertainty of the former is large because of the $\approx$ 25\% parallax error and 0.11\,dex $\log(g)$ error. 

Using 3D extinction maps provided by \citet{gre19}\footnote{http://argonaut.skymaps.info/} towards the direction of \ztf, we find a reddening of $E(g-r)=0.46\pm0.03$ at a distance of $2.3^{+0.7}_{-0.4}$\,kpc, which is somewhat lower than the reddening derived by the SED analysis. Nonetheless, \ztf\, is highly reddened and has a PanSTARRS color of $g-r=0.07$\,mag \citep{cham16}. Such a color puts it in the regime of A- to F-type stars with \teff$\approx10,000$\,K, which is significantly cooler than typical hot subdwarfs. With the 3D map extinction maps corrected magnitude, we find an absolute magnitude of $M_{\rm g}=4.4\pm0.6$\,mag, consistent with a hot subdwarf star \citep{gei19}.


To put constraints on the population origin of \ztf\, we calculated its kinematics. The proper motions of the system is taken from the {\it Gaia} data release 2 catalog (\citealt{gai18}, $\mu_\alpha$cos$(\delta)=-3.326\pm0.167\,\mathrm{mas\,yr^{-1}}$, $\mu_\delta=-5.160\pm0.167\,\mathrm{mas\,yr^{-1}}$). We followed the same approach as described in \citet{kup20} following \citet{ode92} and \citet{pau06}. We use the Galactic potential of \citet{all91} as revised by \citet{irr13}. The orbit was integrated from the present to 3\,Gyrs into the past. \ztf\, moves between a distance of 8.6 to 11.6\,kpc from the Galactic center with an eccentricity of 0.07 within a height of 50\,parsec of the Galactic equator. Exactly as for the other Roche lobe-filling hot subdwarf binary, \ztfold, \ztf\, is a member of the Galactic thin disk population.


\begin{figure*}
    \centering
    \includegraphics[width=0.63\textwidth]{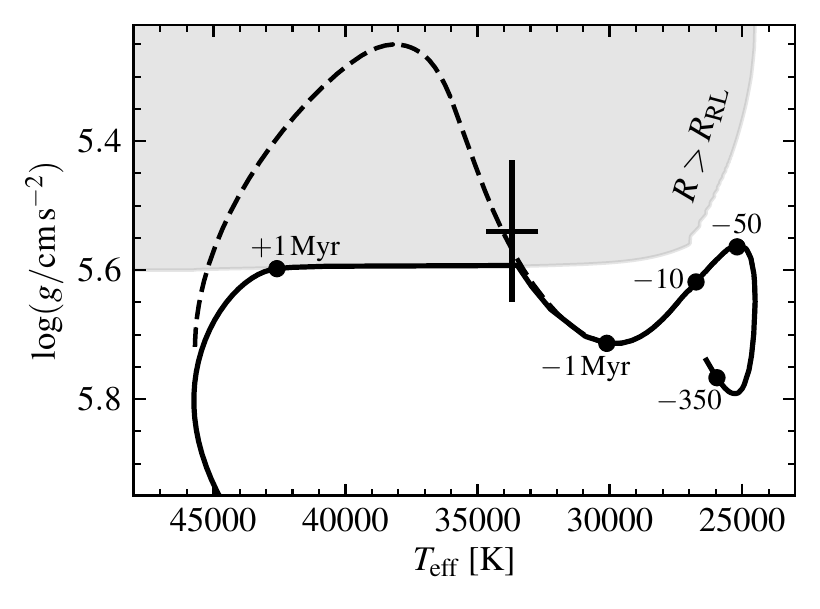}
    \includegraphics[width=0.35\textwidth]{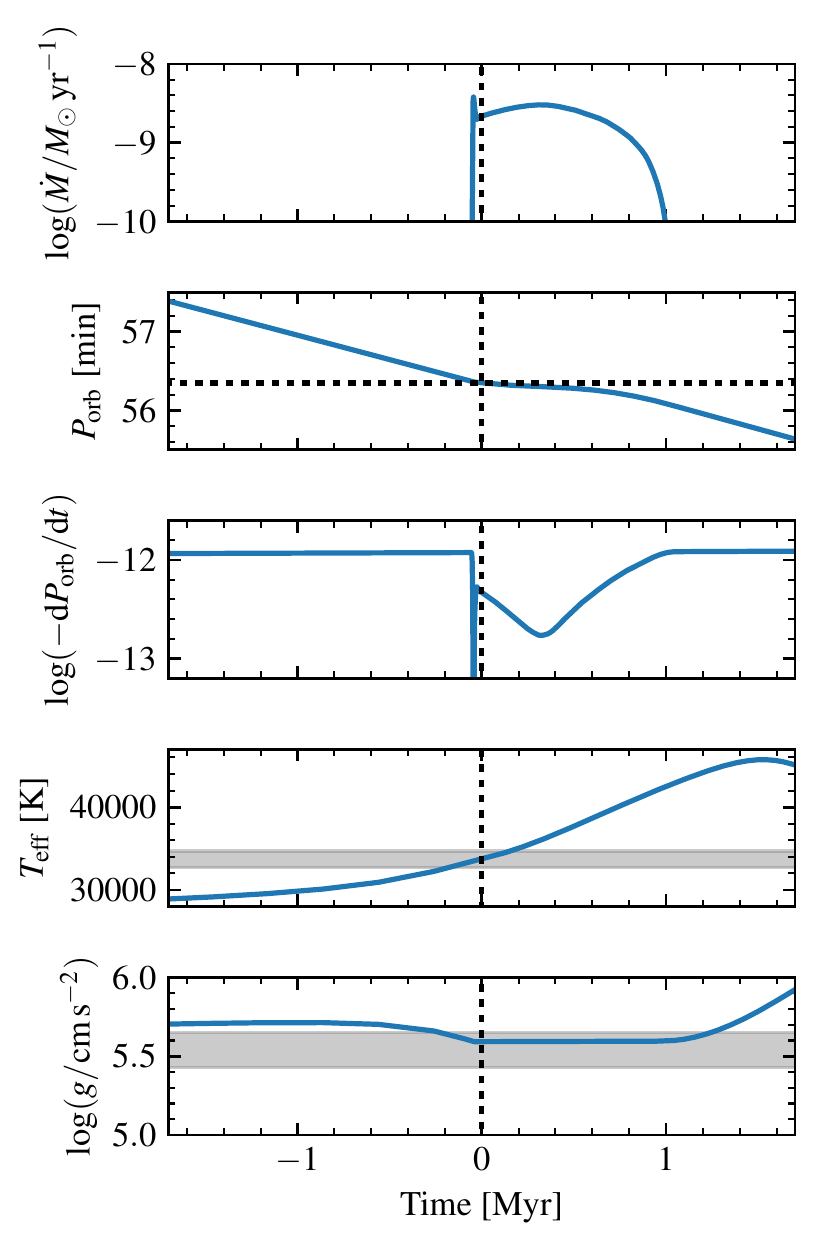}
    \caption{
    {\it Left:} Evolution track for the 0.39\,\msol{} hot subdwarf model. The gray shaded region shows the maximum radius $R$ that the subdwarf can reach before overflowing its Roche lobe radius $R_{\rm RL}$, which shrinks over time due to gravitational wave radiation. The black-dashed curve shows the evolution track for a single-star evolution model of the hot subdwarf that is not confined by the Roche lobe. The black points label ages (in Myr) along the track relative to the present time.
    {\it Right:} Time evolution of the {\tt MESA} binary model through the mass transfer phase leading up to the currently observed state defined by $P_{\rm orb} = 56.35$~minutes at time 0. Gray shaded regions in the lower two panels show the measured values of $T_{\rm eff}$ and $\log(g)$ given in Table~\ref{tab:system}.
    }
    \label{fig:evolution}
\end{figure*}

\section{Discussion}

\subsection{Evolution}
Following the same prescription as described in full detail in Sect. 6.2 in \citet{kup20} we employed {\tt MESA} version 12115 \citep{pax11,pax13,pax15,pax18,pax19} to construct evolutionary models for \ztf. To match the observed properties of \ztf{}, we start with a {\tt MESA} model of a progenitor star with a mass of $2.9$\,\msol{}, which has a main sequence lifetime of 350\,Myr. Once the main-sequence star starts to evolve onto the red giant branch (RGB) He core-burning starts under non-degenerate conditions. We remove most of the hydrogen rich envelope, leaving a He core-burning sdB with \teff$\approx25{,}000$\,K, a core mass of 0.36\,\msol{} and a thin H/He envelope of 0.03\,\msol{}. The envelope is defined by where the H mass fraction is above 0.01. This subdwarf model is then placed in a binary with a $0.65$\,\msol{} WD with an orbital period of $\approx140$\,min. Gravitational wave radiation removes angular momentum from the binary and shortens the period. The period does not shrink far enough during He core-burning, which lasts $\approx350$\,Myr, that the sdB fills its Roche Lobe (see left panel in Fig.\,\ref{fig:evolution}).



Once He core-burning finishes the core contracts and hydrogen burning starts, the radius of the hot subdwarf expands beyond its Roche radius, and mass transfer starts at an orbital period close to the observed orbital period. Mass transfer will continue for $\approx1$\,Myrs at a rate exceeding $10^{-9}\ M_\odot\, {\rm yr}^{-1}$ until hydrogen shell burning is finished and the star contracts to become a carbon-oxygen WD with a thick helium layer and a small residual layer of hydrogen. The high accretion rate will heat the accreting WD significantly \citep{tow03}. Models predict a \teff$\approx50{,}000$\,K for accretion rates of $10^{-9}\ M_\odot\, {\rm yr}^{-1}$ \citep{bur19} consistent with the high black-body temperature of the accretor observed in \ztf. Accretion onto the WD companion at this rate will cause unstable hydrogen ignition after $\approx10^{-4}\ M_\odot$ accumulates, leading to classical novae eruptions \citep{Nomoto82,Nomoto07,Wolf13}. This accretion rate predicts a recurrence time of order $10^5$ years for a total of approximately 10 novae. Our binary model suggests that this system is within the first $\approx10$\% of the 1\,Myrs accretion phase. At the current state the orbit will shrink with $\dot{P}\approx-6\times10^{-13}$s\,s$^{-1}$, which will be detectable after a few years of monitoring. The right panel in Fig.~\ref{fig:evolution} shows the evolution of the donor through this mass transfer phase. After accretion has ceased, the orbit of the system will continue to shrink due to the radiation of gravitational waves and the system will merge in $\approx30$\,Myr. Our models predict that there is a substantial He-layer of $\approx0.05$\,\msol{} left in the former hot subdwarf and the total mass of the system is relatively high ($M_{\rm total}\approx1.1$\,\msol). Recent models predict that such a system explodes as a sub-luminous thermonuclear supernova \citep{Perets19,Zenati19}. If the system avoids a thermonuclear supernova it will merge and could evolve into a rapidly rotating single high-mass carbon-oxygen WD \citep{sai08,cla12,sch19}. 

\subsection{The accretion disc}

Our Gemini spectra show no evidence for any disk lines, which limits the accretion disk contribution of the accretion disk to the overall luminosity to $\leq5\,\%$. Our models predict an accretion rate of $10^{-9}$M$_\odot$yr$^{-1}$. From that we can limit the accretion luminosity to be $\approx1.5$\,L$_\odot$ which is significantly smaller than the luminosity of the sdOB star and in agreement with the absence of any signs of the disk in the optical spectra. 

\citet{riv20} reported the non-detection of X-rays in a 3.6\,ks observation with the Neil Gehrels {\it Swift} Observatory. X-rays from disk-accreting WDs are generally emitted from the boundary layer. We expect that $\approx50$\% of the accretion luminosity is released in the boundary layer. This corresponds to $\approx3\times10^{31}$\,erg\,s$^{-1}$, which is $\approx50\times$ larger than the X-ray upper limit. This suggests an optically-thick boundary layer in \ztf\, with most flux emitted in the extreme UV. This has been observed in SS\,Cyg, where the X-ray luminosity dropped and the extreme UV emission increased during outburst, as the accretion rate increased \citep{whe03}.

\subsection{Implications for a population of Roche lobe-filling hot subdwarfs}
\citet{kup20} presented the first known Roche lobe-filling hot subdwarf binary. \ztf\, is the second known system and the overall evolution appears to be similar to \ztfold. Both systems have $M_{\rm sdOB}<0.47$\,\msol{} hot subdwarf masses and likely evolved from stars with masses $>2$\,\msol\, and therefore are members of a young stellar population consistent with the observed Galactic kinematics. This result is also consistent with predictions by \citet{han03} who found that the most compact hot subdwarf binaries with a compact companion are expected to evolve from main sequence stars with masses $>2$\,\msol. Their study showed that hot subdwarf binaries formed with orbital periods $\lesssim2$\,hrs are dominated by systems with progenitor masses $>2$\,\msol. 


\ztfold\, and \ztf\, are hotter compared to most known hot subdwarfs and can be best explained as donor stars during the shell burning phase. Given that the residual hydrogen shell-burning lifetime of the hot subdwarf is a factor of $\approx10-100$ shorter than its helium core burning lifetime we expect a population $\approx10-100$ larger of systems during core-burning. We suggest here that Galactic reddening could bias discoveries towards preferentially finding Roche lobe-filling systems during shell burning. 


Systems located at low Galactic latitudes will suffer from extinction and reddening. Although the exact extinction is direction dependent, on average the Galactic Plane has an extinction of $\approx1$\,mag per kpc in the optical bands where bluer bands are stronger affected \citep{gon16,gre19}. This leads to a shift of PanSTARRS $g-r$ colors of $\approx0.25-0.3$\,mag per kpc towards redder color leading to a biased sample of hotter or less distant systems in color selected samples. Typically hot subdwarfs have PanSTARRS colors of $g-r\approx-0.35$\,mag \citep{gei19} leading to $g-r\approx0.15-0.25$\,mag for a 2\,mag extinction at a distance of 2\,kpc. These are typical colors for F- to G-type main sequence stars and such objects will be missed in color selected samples for blue stars. Only a reddening-corrected sample, which requires distance estimates, will allow to candidates to be selected in regions with significant extinction. However, particularly in dense Galactic Plane regions, Gaia DR2 data often provide bad astrometry. resulting in many spurious sources. Therefore, hot subdwarfs selected based on absolute magnitudes with reddening corrections are currently done best outside the Galactic Plane with no crowding \citep{gei19}, but misses these hot subdwarfs at low Galactic latitudes that have evolved from a young stellar population and progenitor masses $>2$\,\msol. 

Further discoveries and extended studies using reddening-corrected samples are needed for space density calculations and to understand whether Roche lobe-filling hot subdwarfs are more common amongst shell burning hot subdwarfs rather than He-core burning hot subdwarf stars. 

\section{Conclusions and Summary}
\ztf\, was discovered as a \porb=$56.34785(26)$\,min variable with a light curve shape similar to \ztfold. Follow-up observations show that \ztf\, is the second Roche lobe-filling hot subdwarf binary known today. High signal-to-noise ratio photometry obtained with HiPERCAM in combination with Gemini spectroscopy constrain the system parameters to a mass ratio $q = M_{\rm sdOB}/M_{\rm WD}=0.60\pm0.03$, a mass for the sdOB  $M_{\rm sdOB}=0.41\pm0.04$\,\msol\, and a WD companion mass $M_{\rm WD}=0.68\pm0.05$\,\msol. Additionally, the HiPERCAM light curve revealed a weak ($\approx$1\,\%) eclipse of the hot WD which allowed us to constrain its black-body temperature to $63,000\pm10,000$\,K.

We use the stellar evolution code \texttt{MESA} to compute the evolution of the system and find that the hot subdwarf formed in a common envelope phase at a period of $\approx140$\,min and filled its Roche Lobe close to the observed period when the hot subdwarf expanded during residual hydrogen shell burning. We predict that the binary will transfer mass for $\approx1$\,Myrs and we currently observe the binary in its first 10\,\% of this phase. Our models predict a mass transfer rate of more than $10^{-9}\ M_\odot\, {\rm yr}^{-1}$. Due to its high total mass the system will merge in $\approx 30$\,Myrs and either explode as sub-luminous type Ia supernova \citep{Perets19, Zenati19} or merge and form a massive single WD \citep{1984ApJ...277..355W,sai08,cla12,sch19}.

\ztf\, and \ztfold\, are the only known binaries with a Roche lobe-filling hot subdwarf donor. Both systems have similar properties and we see mass transfer during the short lived hydrogen shell-burning phase compared to the He-core burning phase which lasts significantly longer. The most likely explanation is a selection criteria bias against more typical hot subdwarfs with \teff$\approx25,000$\,K due to severe reddening at low Galactic latitudes. Future studies require reddening corrected samples. We predict that a substantial population of He-core burning Roche lobe-filling hot subdwarf binaries can still be discovered at low Galactic latitudes in young stellar populations.

\acknowledgments

Based on observations obtained with the Samuel Oschin Telescope 48-inch at the Palomar Observatory as part of the Zwicky Transient Facility project. ZTF is supported by the National Science Foundation under Grant No. AST-1440341 and a collaboration including Caltech, IPAC, the Weizmann Institute for Science, the Oskar Klein Center at Stockholm University, the University of Maryland, the University of Washington, Deutsches Elektronen-Synchrotron and Humboldt University, Los Alamos National Laboratories, the TANGO Consortium of Taiwan, the University of Wisconsin at Milwaukee, and Lawrence Berkeley National Laboratories. Operations are conducted by COO, IPAC, and UW.

Based on observations obtained at the international Gemini Observatory, proposal ID GN-2019B-FT-102, a program of NSF’s OIR Lab, which is managed by the Association of Universities for Research in Astronomy (AURA) under a cooperative agreement with the National Science Foundation. on behalf of the Gemini Observatory partnership: the National Science Foundation (United States), National Research Council (Canada), Agencia Nacional de Investigaci\'{o}n y Desarrollo (Chile), Ministerio de Ciencia, Tecnolog\'{i}a e Innovaci\'{o}n (Argentina), Minist\'{e}rio da Ci\^{e}ncia, Tecnologia, Inova\c{c}\~{o}es e Comunica\c{c}\~{o}es (Brazil), and Korea Astronomy and Space Science Institute (Republic of Korea). The authors thank the staff at the Gemini-North observatory for performing the observations in service mode.

Some results presented in this paper are based on observations made with the Shane telescope. Research at Lick Observatory is partially supported by a generous gift from Google.

Based on observations made with the Gran Telescopio Canarias (GTC), installed at the Spanish Observatorio del Roque de los Muchachos of the Instituto de Astrofísica de Canarias, in the island of La Palma.

This research was supported in part by the National Science Foundation through grant ACI-1663688, and at the KITP by grant PHY-1748958. This research benefited from interactions that were funded by the Gordon and Betty Moore Foundation through Grant GBMF5076. 

We acknowledge the use of the Center for Scientific Computing supported by the California NanoSystems Institute and the Materials Research Science and Engineering Center (MRSEC) at UC Santa Barbara through NSF DMR 1720256 and NSF CNS 1725797.

HiPERCAM and VSD are funded by the European Research Council under the European Union’s Seventh Framework Programme (FP/2007-2013) under ERC-2013-ADG Grant Agreement no. 340040 (HiPERCAM).

TRM was supported by a grant from the United Kingdom's Science and Technology Facilities Council. PS acknowledges support from NSF grant AST-1514737. ESP’s research was funded in part by the Gordon and Betty Moore Foundation through Grant GBMF5076. DS acknowledges support by the Deutsche Forschungsgemeinschaft through grant HE1356/70-1.

This work has made use of data from the European Space Agency (ESA) mission {\it Gaia} (\url{https://www.cosmos.esa.int/gaia}), processed by the {\it Gaia} Data Processing and Analysis Consortium (DPAC,
\url{https://www.cosmos.esa.int/web/gaia/dpac/consortium}). Funding for the DPAC has been provided by national institutions, in particular the institutions
participating in the {\it Gaia} Multilateral Agreement.

This work benefited from a workshop held at DARK in July 2019 that was funded by the Danish National Research Foundation (DNRF132).
We thank Josiah Schwab for his efforts in organising this.

\facilities{PO:1.2m (ZTF), Gemini:Gillett (GMOS), Shane (KAST), GTC (HiPERCAM)}

\software{\texttt{Lpipe} \citep{per19}, \texttt{Gatspy} \citep{van15, van15a}, \texttt{FITSB2} \citep{nap04a}, \texttt{LCURVE} \citep{cop10}, \texttt{emcee} \citep{for13}, \texttt{MESA} \citep{pax11,pax13,pax15,pax18,pax19}, \texttt{Matplotlib} \citep{hun07}, \texttt{Astropy} \citep{astpy13, astpy18}, \texttt{Numpy} \citep{numpy}, \texttt{ISIS} \citep{2000ASPC..216..591H}, \texttt{TLUSTY} \citep{hub95}}

\bibliography{refs}{}
\bibliographystyle{aasjournal}

\end{document}